\documentclass[10pt, twoside]{article}
\usepackage{epsf,graphicx,color,amsbsy}
\begin{document}                                                                
                                                                                
\title{\bfseries                                                                
Fossils of turbulence and non-turbulence in the primordial universe:
 the fluid mechanics of dark matter}                  
\vspace{10mm}                                                                  
\author{ Carl H. Gibson}                                                             
\date{}                                                                         
\maketitle                                                                      
\vspace*{-9 mm}                                                                 
{\small                                                                         
\begin{center}                                                                  
Departments of Mechanical and Aerospace Engineering\\
 and Scripps Institution of Oceanography\\                              
University of California at San Diego, La Jolla, CA 92093-0411, USA                                                                                                                       
Contact e-mail:{\texttt cgibson@ucsd.edu, http://www-acs.ucsd.edu/$\sim$
ir118}                                    
\end{center}                                                             
}             
\thispagestyle{empty}
\section{Introduction}                                                          
Was the primordial universe turbulent or non-turbulent soon after the Big Bang? 
How  did the hydrodynamic state of the early universe affect the formation of
structure from gravitational forces, and how did the formation of structure by
gravity affect the hydrodynamic state of the flow?  What can be said about the
dark matter that comprises $99.9 \%$ of the mass of the universe according to
most cosmological models?  Space telescope measurements show answers to these
questions persist literally frozen as fossils of the primordial turbulence
and nonturbulence that controlled structure formation, contrary to standard
cosmology which relies on the erroneous Jeans 1902 linear-inviscid-acoustic
theory and a variety of associated misconceptions (e. g., cold dark matter). When
effects of viscosity, turbulence, and diffusion are included,  vastly different
structure scenarios and  a clear explanation for the dark matter
emerge~\cite{gib96}.  From Gibson's 1996 theory the baryonic (ordinary) dark
matter is comprised of proto-globular-star-cluster (PGC) clumps of hydrogenous
planetoids termed ``primordial fog particles''(PFPs), observed by Schild 1996 as
``rogue planets ... likely to be the missing mass'' of a quasar lensing
galaxy~\cite{sch96}.  The weakly collisional non-baryonic dark matter diffuses
to form outer halos of galaxies and galaxy clusters~\cite{tys95}.
\\

\section{Fluid mechanics of structure formation}  
Before the $1989$ Cosmic Microwave Background Experiment (COBE) satellite, it
was  generally assumed that the fluid universe produced by the hot Big Bang
singularity must be enormously turbulent, and that galaxies were nucleated by
density perturbations produced by this primordial turbulence.  George Gamov
$1954$ suggested galaxies were a form of ``fossil turbulence'', thus coining a
very useful terminology for the description of turbulence remnants in the
stratified ocean and atmosphere, Gibson $1980-1999$.  Other galaxy models based
on turbulence were proposed by von Weizsacker $1951$, Chandrasekhar $1952$, 
Ozernoi and colleagues in $1968-1971$, Oort $1970$, and Silk and Ames $1972$. 
All such  theories were rendered moot by COBE measurements showing temperature
fluctuation values $\delta{T}/T$ of only $10^{-5}$ at $300,000$ years compared
to at least $10^{-2}$ for the plasma if it were turbulent.  At this time, the
opaque plasma of hydrogen and helium had cooled to $3,000$ K and become a
transparent neutral gas, revealing a remarkable  photograph of the universe as it
existed at $10^{13}$ s, with spectral redshift z of $1100$ due to straining of
space at rate $\gamma \approx 1/t$. 
\\
\\
Why was the primordial plasma before $300,000$ years not turbulent? 
Steady inviscid flows are absolutely 
unstable. Turbulence always forms in flows with Reynolds number $Re =
\delta{v} L/\nu$ exceeding $Re_{cr} \approx 100$, where $\nu$ is the kinematic
viscosity of a fluid with velocity differences $\delta v$ on scale $L$,
Landau-Lifshitz 1959.  Thus either $\nu$ at $10^{13}$ s had an
unimaginably large value of $9\times10^{27}$ m$^2$ s$^{-1}$ at horizon scales
$L_H = ct$ with light speed velocity differences $c$, or else gravitational
structures formed in the plasma at earlier times and viscosity plus buoyancy
forces of the structures prevented strong turbulence.\\

\section{Fossils of first structure (proto-supervoids)}
The power spectrum of temperature fluctuations $\delta T$  measured by 
COBE peaks at a length $3 \times 10^{20}$ m which is only $1/10$ the horizon
scale ct, suggesting the first structure formed earlier at $10^{12}$ s
($30,000$ years).  The photon viscosity of the plasma $\nu=c/n\sigma_{\tau}$ was
$4\times{10^{26}} $ m$^{2}$ s$^{-1}$ then, with free electron number
density
$n=10^{10}$ m$^{-3}$ and $\sigma_{\tau}$ the Thomson cross section for Compton
scattering.  The baryon density $\rho$  was
$3\times10^{-17}$ kg m$^{-3}$, which matches the density of present 
globular-star-clusters as a fossil of the weak turbulence at this time of first
structure.    The fragmentation mass
$\rho(ct)^{3}$ of $10^{46}$ kg matches the observed mass of superclusters of
galaxies, the largest structures of the universe. Because $Re \approx
Re_{crit}$, the horizon scale
$ct=3\times{10^{20}}$ m matches the Schwarz viscous scale $L_{SV} =
(\gamma\nu/\rho{G})^{1/2}$ at which viscous forces
$F_{V}=\rho\gamma{L}^{2}$ equal gravitational forces $F_{G}=\rho^{2}GL^{4}$, and
also the Schwarz turbulence scale $L_{ST} = \varepsilon^{1/2}/(\rho
G)^{3/4}$ at which inertial-vortex forces
$F_I = \rho \varepsilon^{2/3} L^{8/3}$ equal $F_{G}$, where $\varepsilon$ is the
viscous dissipation rate~\cite{gib96}.  Further fragmentation to proto-galaxy
scales is predicted in this scenario, with the   nonbaryonic dark matter
diffusing to fill the voids between constant density proto-supercluster to
proto-galaxy structures for scales smaller than the diffusive Schwarz scale
$L_{SD}=(D^{2}/\rho{G})^{1/4}$, where $D$ is the diffusivity of the nonbaryonic
dark matter~\cite{gib96}.  Fragmentation of the nonbaryonic material to form
superhalos implies
$D=10^{28}$ m$^{2}$ s$^{-1}$, from observation of present superhalo sizes
$L_{SD}$ and densities
$\rho$~\cite{tys95}, trillions of times larger than $D$ for H-He gas with the
same $\rho$. 
\\

\section{Fossils of the first condensation (as ``fog'')}
Photon decoupling dramatically reduced viscosity values to $\nu =
3\times{10^{12}}$ m$^{2}$ s$^{-1}$ in the primordial gas of the nonturbulent
$10^{20}$ m size proto-galaxies, with $\gamma = 10^{-13}$ s$^{-1}$ and $\rho =
10^{-17}$ kg m$^{-3}$, giving a PFP fragmentation mass range $M_{SV} \approx
M_{ST}
\approx  10^{23}-10^{25}$ kg, the mass of a small planet.  Pressure decreases in
voids during fragmentation as the density decreases, to maintain constant
temperature from the perfect gas law
$T=p/\rho{R}$, where $R$ is the gas constant, for scales  smaller than the
acoustic scale $L_{J} = V_{S}/(\rho{G})^{1/2}$ of Jeans $1902$, where $V_{S}$ is
the sound speed.  However, the pressure cannot propagate fast enough in
voids larger than $L_J$ so they cool.  Hence radiation from the warmer
surroundings can heat such large voids, increasing their pressure and
accelerating the void formation, causing a fragmentation within
proto-galaxies at the Jeans mass of $10^{35}$ kg, the mass of
globular-star-clusters.  These proto-globular-cluster (PGC) clumps of PFPs
provide the materials of construction for everything else to follow, from stars
to people. Leftover PGCs and PFPs thus comprise present galactic dark matter
inner halos which typically have expanded to about
$10^{21}$ m (30 kpc) of the core and exceed the luminous (star) mass by factors
of
$10-30$.
\\

\section{Observations}
Observations of quasar image twinkling frequencies reveal that the point mass
objects which dominate the mass of the lens galaxy are not stars, but ``rogue
planets... likely to be the missing mass'', Schild $1996$, independently
confirming this prediction of Gibson $1996$.  Other evidence of the predicted
primordial fog particles (PFPs) is shown in Hubble Space Telescope photographs,
such as thousands of 
$10^{25}$ kg ``cometary globules'' in the halo of the Helix planetary nebula and
possibly like numbers in the Eskimo planetary nebula halo.  These dying stars are
very hot ($100,000$ K versus $6,000$ K normal) so that many 
PFPs  nearby can be brought out of cold storage by evaporation to produce the
$10^{13}$ m protective cocoons that make them visible to the HST at $10^{19}$ m
distances.
\\ 
 
\section{Summary and conclusions}
The Figure summarizes the evolution of structure and turbulence in the early
universe, as inferred from the present  nonlinear fluid mechanical theory. It is
very different, very early, and very gentle compared to the standard model, where
structure formation in baryonic matter is forbidden in the plasma epoch because
$L_J$ is larger than $L_H = ct$ and 
galaxies collapse at 140 million years (redshift
z=20) producing $10^{36}$ kg Population III superstars
that explode and re-ionize the universe to explain the missing gas (sequestered
in PFPs).  No such stars, no galaxy collapse, and no re-ionization occurs in the
present theory.  To produce the structure observed today, the concept ``cold
dark matter'' (CDM) was invented; that is, a hypothetical non-baryonic fluid of
``cold'' (low speed) collisionless particles with adjustable $L_J$ small enough
to produce gravitational potential wells to drive galaxy collapse.   Cold dark
matter is unnecessary in the present theory. Even if it exists it would not
behave as required by the standard model.  Its necessarily small collision cross
section requires
$L_{SD} \gg L_J$ so it would diffuse out of its own well, without fragmentation
if $L_{SD} \gg L_H$. The immediate formation of ``primordial fog
particles'' from all the neutral gas of the universe emerging from the plasma
epoch permits their gradual accretion to form the observed small ancient stars in
dense globular-star-clusters known to be only slightly younger than the
universe.  These could never form in the intense turbulence of galaxy collapse in
the standard model because
$L_{ST}$ scales would be too large. 
\\

\begin{figure}[!h]                                                             
   \vspace{1mm}                                                                
   \begin{center}                                                               
   \includegraphics[width= 0.8 \linewidth]{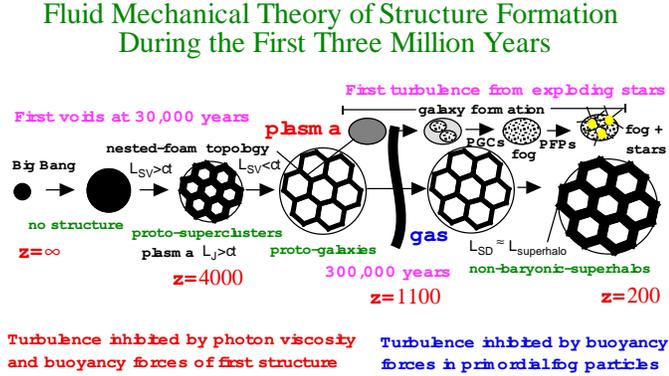}                    
   \end{center}                                                                 

\vskip-2.8in
                                                         
   \caption{Evolution of structure and turbulence in the early universe}                                                
   \end{figure}

\end{document}